%% file: main.tex
\documentclass[sigconf, nonacm]{acmart}

\usepackage{a4wide}
\usepackage{mathptmx}
\usepackage{multirow}
\usepackage{graphicx}
\usepackage{subcaption}
\usepackage{amsfonts}
\usepackage{color}
\usepackage[linesnumbered,ruled]{algorithm2e}
\usepackage{listings}
\AtBeginDocument{%
  }

\setcopyright{acmlicensed}
\copyrightyear{2024}
\acmYear{2024}
\acmDOI{XXXXXXX.XXXXXXX}

\begin{document}

\title{Supporting Cross-language Cross-project Bug Localization Using Pre-trained Language Models}

\author{Mahinthan Chandramohan}
\affiliation{%
  \institution{Oracle Labs}
  \city{Brisbane}
  \country{Australia}
}

\author{Dai Quoc Nguyen}
\affiliation{%
  \institution{Oracle Labs}
  \city{Brisbane}
  \country{Australia}
}

\author{Padmanabhan Krishnan}
\affiliation{%
  \institution{Oracle Labs}
  \city{Brisbane}
  \country{Australia}
}

\author{Jovan Jancic}
\affiliation{%
  \institution{Oracle Labs}
  \city{Vienna}
  \country{Austria}
}

\begin{abstract}
Automatically locating a bug within a large codebase remains a significant challenge for developers. Existing techniques often struggle with generalizability and deployment due to their reliance on application-specific data and large model sizes.

This paper proposes a novel pre-trained language model (PLM) based technique for bug localization that transcends project and language boundaries. Our approach leverages contrastive learning to enhance the representation of bug reports and source code. It then utilizes a novel ranking approach that combines commit messages and code segments. Additionally, we introduce a knowledge distillation technique that reduces model size for practical deployment without compromising performance.

This paper presents several key benefits. By incorporating code segment and commit message analysis alongside traditional file-level examination, our technique achieves better bug localization accuracy. Furthermore, our model excels at generalizability –  trained on code from various projects and languages, it can effectively identify bugs in unseen codebases. To address computational limitations, we propose a CPU-compatible solution.

Our approach demonstrates great performance. On average, there is a 73-80\% chance of identifying at least one correct bug location within the top 10 results for cross-language scenarios, and a 55-87\% chance for cross-projects scenarios. Notably, this is achieved without fine-tuning the model on the specific datasets evaluated. The proposed fine-tuning and ranking algorithms significantly outperform off-the-shelf PLMs for bug localization. Finally, our distillation technique achieves performance comparable to the original model while offering a 4x reduction in embedding generation time. In essence, this work presents a highly effective, generalizable, and efficient bug localization technique with the potential to real-world deployment.

\end{abstract}



\maketitle

\section{Introduction}

Addressing customer reported bugs promptly is crucial for any development team.
These bugs could manifest as unexpected behavior, crashes, or incorrect outputs, ultimately hindering software functionality and user experience.
However, locating the exact location of a bug within the vast codebase, a process known as bug localization, can be a time-consuming and laborious task.
Despite many years of research, bug localization remains a challenging
problem \cite{2016WGL-TSE}.
Developers often spend a significant portion of their time manually sifting
through lines of code, analyzing logs, and debugging to identify the root cause
of the issue \cite{2014MHM-JSS}. 

Bug localization within modern software systems presents a significant challenge due to several factors~\citep{du2023pre, ciborowska2022fast, wen2016locus}. The sheer size and complexity of codebases, often encompassing millions of lines of code, makes locating the specific buggy segment a daunting task. Additionally, developer expertise can be limited by the domain-specific nature of the software or their unfamiliarity with particular code sections. Furthermore, the effectiveness of bug localization can be hampered by the quality of bug reports. Incomplete or ambiguous reports submitted by users or testers may lack the necessary details or clarity to accurately guide developers towards the root cause. Finally, even well-described bug reports may not exhibit a clear correlation between the bug's manifestation and its actual location within the source code.

Therefore, automated bug localization techniques have emerged as a vital area of research to address these challenges`\citep{ciborowska2022online,du2023pre,wen2016locus,ciborowska2022fast}. These techniques aim to automate, at least partially, the process of pinpointing the source code responsible for a bug
report. By leveraging various methodologies, such as machine learning,
information retrieval, and code analysis, automated bug localization aims to
reduce the time and effort required for developers to locate bugs and improve the efficiency of the bug fixing process.

In recent times, pre-trained language models (PLMs) (e.g., CodeBERT
\cite{codebert2020}, UniXcoder \cite{guo2022unixcoder}) have become an important
trend for code representation learning and produce significant improvement for
programming language prediction and classification tasks (e.g., code clone
detection \cite{white2016deep} and bug localization \cite{zhou2012should}).
State-of-the-art bug localization techniques proposed in the literature
are application and language specific. Hence, such techniques do not generalize well to
support cross-application, cross-language use cases. Further, pre-trained
language model based techniques lack practical deployability. This is because
the deploying pre-trained LMs having hundred million parameters requires
devices to have appropriate hardware resources such as GPUs which are expensive
and hence not available to most of the developers.

However, as pre-trained large language models have recently become an essential trend for code representation learning, we examine pre-trained large language models for cross-project and cross-language bug localization. In this paper, we propose a novel technique to enhance learning better representations of bug reports and source code files, by integrating supervised contrastive learning during fine-tuning of pre-trained large language model and  utilizing a simple yet effective method for sampling better negative samples. We also propose a novel ranking technique that leverages code segments and commit message analysis to effectively locate source files, code segments and commits messages relevant to the bug. Finally, to enable practical deployability, we also propose a novel knowledge distillation technique to reduce the model size to make it CPU-compatible without compromising the performance. 

The existing approaches \cite{lee2018bench4bl,chakraborty2023rlocator} consider file-level analysis for bug localization by computing the cosine similarity between the given bug report and the whole project file. However, pre-trained language models has a limited input token size; hence, the models may ignore partial code that is more important and more relevant to the given bug report. To deal with this issue, we consider code segment-level analysis by splitting the given source code file into consecutive code segments, thus enhancing searching relevant files. Further, existing techniques such as FBL-BERT \cite{ciborowska2022fast} focuses on commit hunk-level analysis to find relevant commit hunks for a given bug report. Our goal is different from FBL-BERT, wherein we consider relevant commit messages for a given bug report to locate relevant source files and code segments. To this end, we combine strong points of both code segment-level and commit message-level analyses to increase bug localization performance in general.

On the other hand, recent bug localization techniques leveraging pre-
trained language models require powerful GPUs~\citep{ciborowska2022fast, du2023pre}, which are expensive and not widely available. Knowledge distillation techniques have been proposed to address this challenge~\citep{2020mobilebert, jiao2020tinybert, wang2021minilmv2, wang2023distill}, but they have shortcomings. For example, existing approaches (e.g., TinyBERT \cite{jiao2020tinybert}, and MobileBERT \cite{2020mobilebert}) often consider layer-to-layer distillation  using uniform layer mapping. However, \cite{wang2021minilmv2} show that layer-to-layer distillation mapping leads to over-fit training samples. In addition, in existing techniques, e.g., \cite{wang2021minilmv2}, student model weights are randomly initialized, which is not desired. Further, in existing approaches  \cite{jiao2020tinybert,wang2023distill} consider the distillation loss functions only when fine-tuning the student model for task-specific distillation. This restricts the student model to only imitating the behavior of the teacher model, leading to inferior performance. 

In this paper we describe our solution to the bug localization problem. That is, our technique helps the developers locate the source files, code segments and commits that are relevant to fix the reported bug \cite{zhou2012should} because identifying the relevant buggy methods in large code repositories is time-consuming. We further propose a novel knowledge distillation technique to generate a CPU-compatible, scaled-down version of the model. \\

In summary, contributions of our paper are as follows: 
\begin{itemize}
\item We developed a novel bug localization technique that identifies relevant files, code segments, and commit messages across projects and languages
\item We present a contrastive learning based pre-trained language model fine-tuning technique, where we propose a simple yet effective negative sample generation technique. 
\item We present a novel knowledge distillation technique to reduce the model
size without compromising the model performance.
\item We show how our techniques generalizes for different applications written
in different languages.
\end{itemize}

\input{motivation}

\input{methodology}

\input{implementation}

\input{exprDesign}
\input{results}
\input{threats}
\input{relatedWork}
\input{conclusion}


\bibliographystyle{ACM-Reference-Format}
\bibliography{references}



%
%

\end{document}

%% file: motivation.tex
\section{Motivation} \label{sec:motivation}

\begin{table}[]
\begin{tabular}{|l|r|r|r|r|r|r|}
\hline

\multicolumn{1}{|c|}{Project} & \multicolumn{1}{c|}{\begin{tabular}[c]{@{}c@{}}\# \\ Bugs\end{tabular}} & \multicolumn{1}{c|}{\begin{tabular}[c]{@{}c@{}}Bug ind. \\ Commit\end{tabular}} & \multicolumn{1}{c|}{\begin{tabular}[c]{@{}c@{}}Linked\\ Bugs\end{tabular}} & \multicolumn{1}{c|}{\begin{tabular}[c]{@{}c@{}}Aff.\\ Files\end{tabular}} & \multicolumn{1}{c|}{\begin{tabular}[c]{@{}c@{}} \# \\ Add.\end{tabular}} & \multicolumn{1}{c|}{\begin{tabular}[c]{@{}c@{}} \# \\ Del.\end{tabular}} \\ \hline

AspectJ & 200     & d842c4f                                                    & 75 (38\%)                                             & 1,709                                                    & 332,365     & 0           \\ \hline
JDT     & 94      & c2e080f                                                    & 32 (34\%)                                             & 761                                                      & 206,903     & 206,903     \\ \hline
PDE     & 60      & 7523f22                                                    & 14 (23\%)                                             & 923                                                      & 24,761      & 34,097      \\ \hline
SWT     & 90      & 281dc95                                                    & 48 (53\%)                                             & 678                                                      & 205,790     & 205,790     \\ \hline
Tomcat  & 193     & eae5441                                                    & 63 (33\%)                                             & 1,008                                                     & 273,329     & 273,329     \\ \hline
ZXing   & 20      & 7726e8f                                                    & 5 (25\%)                                              & 58                                                       & 5970        & 0           \\ \hline
\end{tabular}
\caption{Summary of projects and their most dominant bug inducing commit found in the changeset-based bug localization dataset as used in~\cite{ciborowska2022fast} and~\cite{du2023pre}}
\label{tab:dataset_motivation}
\end{table}

One of the biggest limitations of existing bug localization research, whether based on machine learning~\citep{ciborowska2022fast,du2023pre} or information retrieval (IR)~\citep{wen2016locus}, is the requirement for training or fine-tuning on a specific dataset. Unfortunately, such training data is not always readily available for each project where bug localization is needed. This lack of generalizability across projects and languages hinders the practical application of existing techniques in real-world settings.

Furthermore, while recently proposed changeset-based bug localization techniques appear promising in theory, their real-world
effectiveness remains questionable. This is primarily due to how developers actually modify codebases. Changesets are ideally small, but this is often not the case in practice. Table~\ref{tab:dataset_motivation} (summarizing projects and dominant bug-inducing commits from the dataset used by techniques proposed in ~\citep{ciborowska2022fast,du2023pre}) exemplifies this. As the table shows, each project has at least one commit linked to a significant portion of the bugs (e.g., commit  \texttt{281dc95} in SWT is linked to 53\% of the total bugs). Manual inspection reveals that these commits are frequently related to refactoring, formatting, or migration activities (e.g., \texttt{SVN} to \texttt{Git}) and involve large code hunks (mostly files). Identifying the actual bug location within these large hunks is impractical, requiring developers to invest significant time and effort.

Additionally, recent bug localization techniques leveraging pre-trained language models~\citep{du2023pre}  necessitate powerful GPUs (e.g., NVIDIA Tesla A100). However, GPUs are expensive and not universally available within product teams. This may limit the widespread adoption of such techniques. There is a need to develop these sophisticated machine learning models to run efficiently on CPUs, enabling broader use.

To address these limitations, this work proposes a bug localization technique that aims to identify files, code segments, and commit messages relevant to the bug across projects and languages, while remaining CPU-compatible.

%% file: methodology.tex
\section{Methodology}

In this section, we first discuss the high-level overview of the proposed bug
localization framework followed by fine-tuning of pre-trained language model for
bug localization and conclude with the our knowledge distillation technique that
enables developers to use our models on typical machines that do not have
specialist hardware like GPUs.

\subsection{Bug-localization Framework}
\begin{figure}[t]
    \centering
        \centering
        \includegraphics[width=0.4\textwidth]{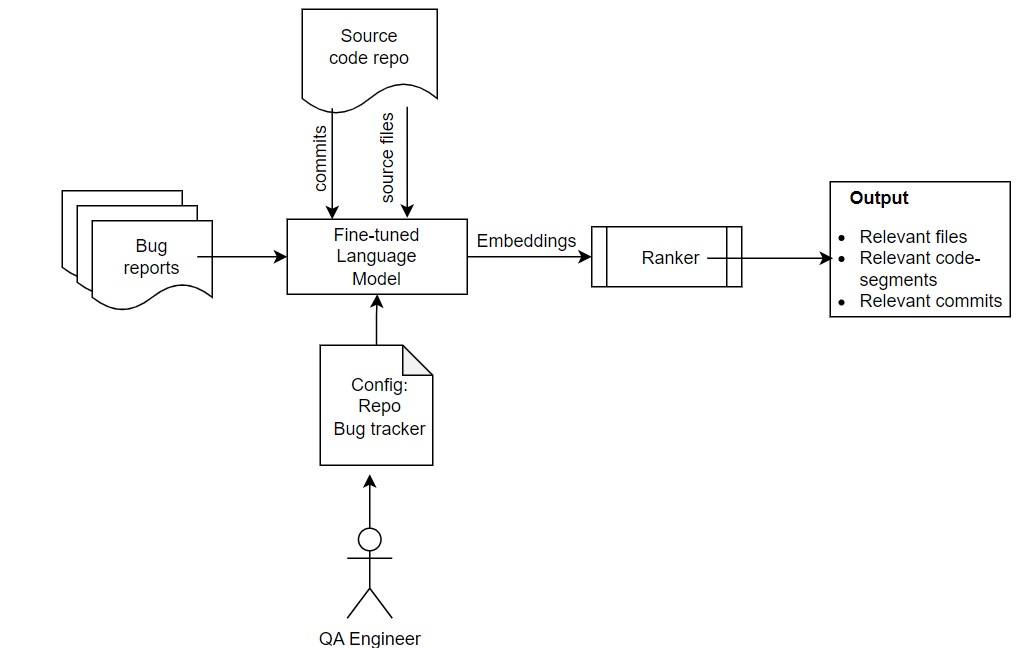} 
        \caption{High-level bug-localization framework}
        \label{fig:localization_architecture}
\end{figure}

The high-level bug-localization architecture is shown in
Figure~\ref{fig:localization_architecture}. The first step is granting our
system access
to the source code repository and bug-tracking system (e.g., JIRA, GitHub
issues, etc.) via a configuration file. This enables embeddings for the source
files and commits and previously reported bugs to be generated. This can be done
``offline'' and periodically to keep the embeddings current. Algorithm~\ref{alg:embedding_generation} shows the steps involved in embedding generation from source files and commit messages. These steps are repeated for new/updated files and commit messages and the embedding store is kept up-to-date.
When a new bug report is filed, embeddings for the bug
report is generated. Both these steps use the fine-tuned language model. The
generated 
embeddings are passed through the ranker module to identify source files, code
segments and commits that are relevant to the new bug report.

\begin{algorithm}[h]
\DontPrintSemicolon
\SetAlgoVlined

\textbf{Input}: Project source files: $\mathcal{F} = \{f_1, f_2, ..., f_m\}$, Commit messages: $\mathcal{C} = \{\left(cm_1, \mathsf{F}_1\right), \left(cm_2, \mathsf{F}_2)\right), ..., \left(cm_n, \mathsf{F}_n\right) | \mathsf{F}_i \subset \mathcal{F}, i = 1, 2, ..., n \}$, and model parameters \textbf{$\theta$}.

\textbf{Output}: Embedding $\mathsf{E_{cs}}$ and $\mathsf{E_{cm}}$, File mapping $\mathsf{C_{cs}}$ and $\mathsf{C_{cm}}$

$\mathsf{E_{cs}} \leftarrow \{\}$

$\mathsf{C_{cs}} \leftarrow \{\}$

\For{$f \in \mathcal{F}$}{

	Split $f$ into consecutive code segments of length of 512 tokenized tokens
	
	\For{$\mathsf{each\ code\ segment}$ cs}{

        $cs_p$ $\leftarrow$ add\_prefix($cs$)
	
		$cs'$ $\leftarrow$ get\_embedding($cs_p$, \textbf{$\theta$})
		
		$\mathsf{E_{cs}} \leftarrow \mathsf{E_{cs}} \cup \{cs, cs'\}$
		
		$\mathsf{C_{cs}} \leftarrow \mathsf{C_{cs}} \cup \{cs , f\}$ \tcp*{code segments to file mapping}
	}
}
$\mathsf{E_{cm}} \leftarrow \{\}$

$\mathsf{C_{cm}} \leftarrow \{\}$

\For{$\left(cm, \mathsf{F}\right) \in \mathcal{C}$}{

    \textbf{e} $\leftarrow$ get\_embedding($cm$, \textbf{$\theta$})
        
    $\mathsf{E_{cm}} \leftarrow \mathsf{E_{cm}} \cup \{\textbf{e}\}$
    
    $\mathsf{C_{cm}} \leftarrow \mathsf{C_{cm}}  \cup \{cm , f\}$ \tcp*{commits to file mapping}
}

\caption{Embedding generation}
\label{alg:embedding_generation}
\end{algorithm}

In the bug-localization framework, we propose a novel ranking technique that operates on code segments and commit messages to locate bugs at various granularity levels. We first propose a code segment-level ranking, wherein we split the source code file into consecutive code segments of length of 512 tokenized tokens with adding a prefix of package and class names at the beginning of each code segment. We then leverage the model to compute the cosine similarity scores between the embedding of the given bug report and the embeddings of the code segments to obtain ranked corresponding files to locate the first relevant file to the bug report. Algorithm~\ref{alg:code_fragment_level_evaluation} summarises the steps involved in code segment ranking.

\begin{algorithm}[h]
\DontPrintSemicolon
\SetAlgoVlined

\textbf{Input}: Embedding $\mathsf{E_{cs}}$, File mapping $\mathsf{C_{cs}}$, input bug reports: $\mathcal{I} = \{(r_1), (r_2), ..., (r_l)\}$, an integer numeric $k$, and model parameters \textbf{$\theta$}.

\textbf{Output}: $ranked\_codesegment\_lists$, $ranked\_file\_lists_{cs}$

$\mathsf{R} \leftarrow \{\}$

\For{$(r) \in \mathcal{I}$}{

	$r'$ $\leftarrow$ get\_embedding($r$, \textbf{$\theta$})
	
	$\mathsf{R} \leftarrow \mathsf{R} \cup \{(r, r')\}$
}
$ranked\_file\_list_{cs} \leftarrow \{\}$

$ranked\_codesegment\_list \leftarrow \{\}$

\For{$(r, r') \in \mathsf{R}$}{
	
	\textbf{s} $\leftarrow$ cosine\_similarity($r'$, $\mathsf{E_{cs}}$) 
	
	$ranked\_list \leftarrow$ descending\_order(\textbf{s}, $\mathsf{C_{cs}}$)

    $file\_list \leftarrow \{\}$
	
    \For{$\left(cs, score\right) \in ranked\_list$[:k]}{
	
		$ranked\_codesegment\_list \leftarrow ranked\_codesegment\_list \cup \{\left(r, \{cs, score\}\right)\}$
  
        $\mathsf{F}$ $\leftarrow$ get\_files($cs$, $\mathsf{C_{cs}}$)
  
        \For{$file \in \mathsf{F}$} {
		
			$file\_list \leftarrow file\_list \cup \{file\}$
		}
	}
	$common\_files \leftarrow$ counter\_most\_common($file\_list$)

    $ranked\_file\_list_{cs} \leftarrow ranked\_file\_list_{cs} \cup \{\left(r, \{common\_files\}\right)\}$
} 

\caption{Code segment ranking}
\label{alg:code_fragment_level_evaluation}
\end{algorithm}

We also observe that the bug reports can be relevant to commit messages. Thus, we consider a commit message-level ranking, wherein we utilize the model to compute the cosine similarity scores between the embedding of the given bug report and the embeddings of the commit messages to obtained top $k$ ranked commits. We then extract the obtained commits' associated files and rank these files according to file occurrence to locate the first relevant file to the bug report. Algorithm~\ref{alg:commit_messages_level_evaluation} depicts the steps involved in commit message ranking.

\begin{algorithm}[h]
\DontPrintSemicolon
\SetAlgoVlined

\textbf{Input}:  Embedding $\mathsf{E_{cm}}$, File mapping $\mathsf{C_{cm}}$, input bug reports: $\mathcal{I} = \{(r_1), (r_2), ..., (r_l)\}$, an integer numeric $k$, and model parameters \textbf{$\theta$}.

\textbf{Output}: $ranked\_commit\_lists$, $ranked\_file\_lists_{cm}$

$\mathsf{R} \leftarrow \{\}$

\For{$(r) \in \mathcal{I}$}{

	$r'$ $\leftarrow$ get\_embedding($r$, \textbf{$\theta$})
	
	$\mathsf{R} \leftarrow \mathsf{R} \cup \{(r, r')\}$
}

$ranked\_file\_list_{cm} \leftarrow \{\}$

$ranked\_commit\_list \leftarrow \{\}$

\For{$(r, r') \in \mathsf{R}$}{
	
	\textbf{s} $\leftarrow$ cosine\_similarity($r'$, $\mathsf{E_{cm}}$) 
	
	$ranked\_lists \leftarrow$ descending\_order(\textbf{s}, $\mathsf{C_{cm}}$)

	$file\_list \leftarrow \{\}$
	
    \For{$\left(cm, score\right) \in ranked\_list$[:k]}{
	
		$ranked\_commit\_list \leftarrow ranked\_commit\_list \cup \{\left(r, \{cm, score\}\right)\}$
  
        $\mathsf{F}$ $\leftarrow$ get\_files($cm$, $\mathsf{C_{cm}}$)
  
        \For{$file \in \mathsf{F}$} {
		
			$file\_list \leftarrow file\_list \cup \{file\}$
		}
	}
	$common\_files \leftarrow$ counter\_most\_common($file\_list$)

    $ranked\_file\_list_{cm} \leftarrow ranked\_file\_list_{cm} \cup \{\left(r, \{common\_files\}\right)\}$
} 

\caption{Commit message ranking}
\label{alg:commit_messages_level_evaluation}
\end{algorithm}

\begin{figure}[!ht]
    \centering
    \begin{minipage}{0.5\textwidth}
        \centering
        \includegraphics[width=1\textwidth]{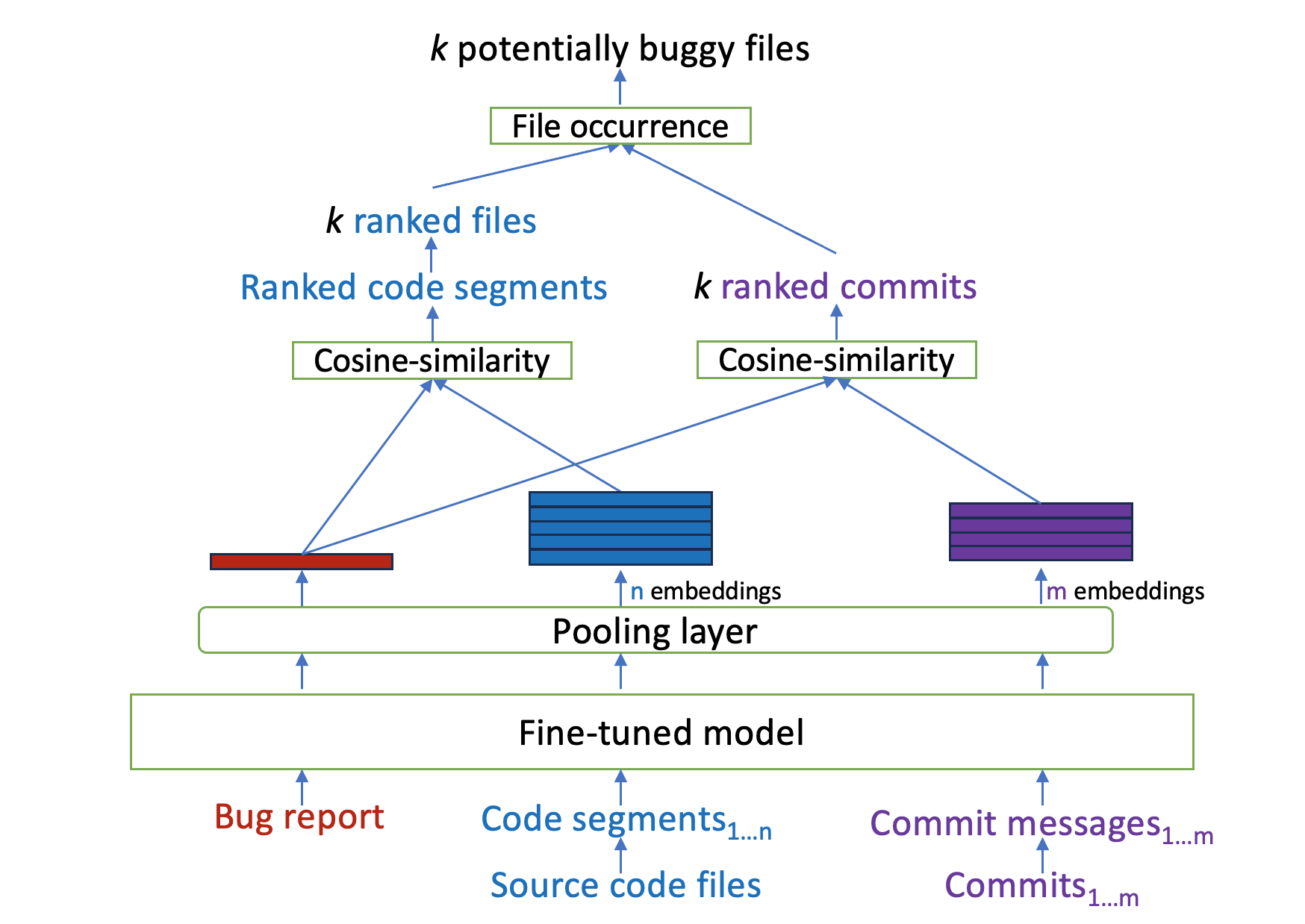} 
        \caption{Proposed ranking technique}
        \label{fig:overview_proposed_evaluation}
    \end{minipage}
\end{figure}

The proposed ranking technique is a combination of code segment and commit message rankings. As summarized in Algorithm~\ref{alg:combined_evaluation}, we consider a list of top $k$ ranked files with their similarity scores from the code segment-level analysis and a list of top $k$ ranked commits with their similarity scores from commit message-level analysis. 
We then combine two list into a single list and sort the list according to descending order of similarity scores to extract top $k$ ranked items (e.g., each item can be a file or a commit).
After that, we extract all relevant files from the top $k$ ranked items and re-rank the files according to file occurrence to locate the relevant file to the bug report. We further get commit messages and code segments associated with those files. The high-level overview of the proposed ranking is shown in Figure~\ref{fig:overview_proposed_evaluation}.

\begin{algorithm}[!ht]
\DontPrintSemicolon
\SetAlgoVlined

\textbf{Input}: Embedding $\mathsf{E_{cs}}$ and $\mathsf{E_{cm}}$, File mapping $\mathsf{C_{cs}}$ and $\mathsf{C_{cm}}$, input bug reports: $\mathcal{I} = \{(r_1), (r_2), ..., (r_l)\}$, an integer numeric $k$, and model parameters \textbf{$\theta$}

\textbf{Output}: $ranked\_file\_cs\_cm\_list$

$ranked\_file\_list_{cs} \leftarrow $ codesegment\_ranking($\mathsf{E_{cs}}$, $\mathsf{C_{cs}}$, $k$, \textbf{$\theta$}) \tcp*{Algorithm \ref{alg:code_fragment_level_evaluation}}

$ranked\_file\_list_{cm} \leftarrow $ commit\_ranking($\mathsf{E_{cm}}$, $\mathsf{C_{cm}}$, $k$, \textbf{$\theta$}) \tcp*{Algorithm \ref{alg:commit_messages_level_evaluation}}

\For{$(r) \in \mathcal{I}$}{

	$combined\_files \leftarrow$ $ranked\_file\_lists_{cs}$[$r$] $\cup$  $ranked\_file\_lists_{cm}$[$r$]

	$common\_list \leftarrow$ counter\_most\_common($combined\_files$)

    \For{$file \in common\_list$[:k]} {
		
	    $cs$ 	$\leftarrow$ get\_codesegments($file$, $\mathsf{C_{cs}}$)	

        $cm$ 	$\leftarrow$ get\_commits($file$, $\mathsf{C_{cm}}$)	

        $ranked\_file\_cs\_cm\_list \leftarrow ranked\_file\_cs\_cm\_list \cup \{\left(r, \{file\}, \{cs\}, \{cm\}\right)\}$
		
    }
} 

\caption{Proposed ranking algorithm}
\label{alg:combined_evaluation}
\end{algorithm}

\subsection{Fine-tuning Pre-trained Language Model} \label{sec:fine-tuning}
\begin{figure}[!ht]
    \centering
    \begin{minipage}{0.4\textwidth}
        \centering
        \includegraphics[width=0.9\textwidth]{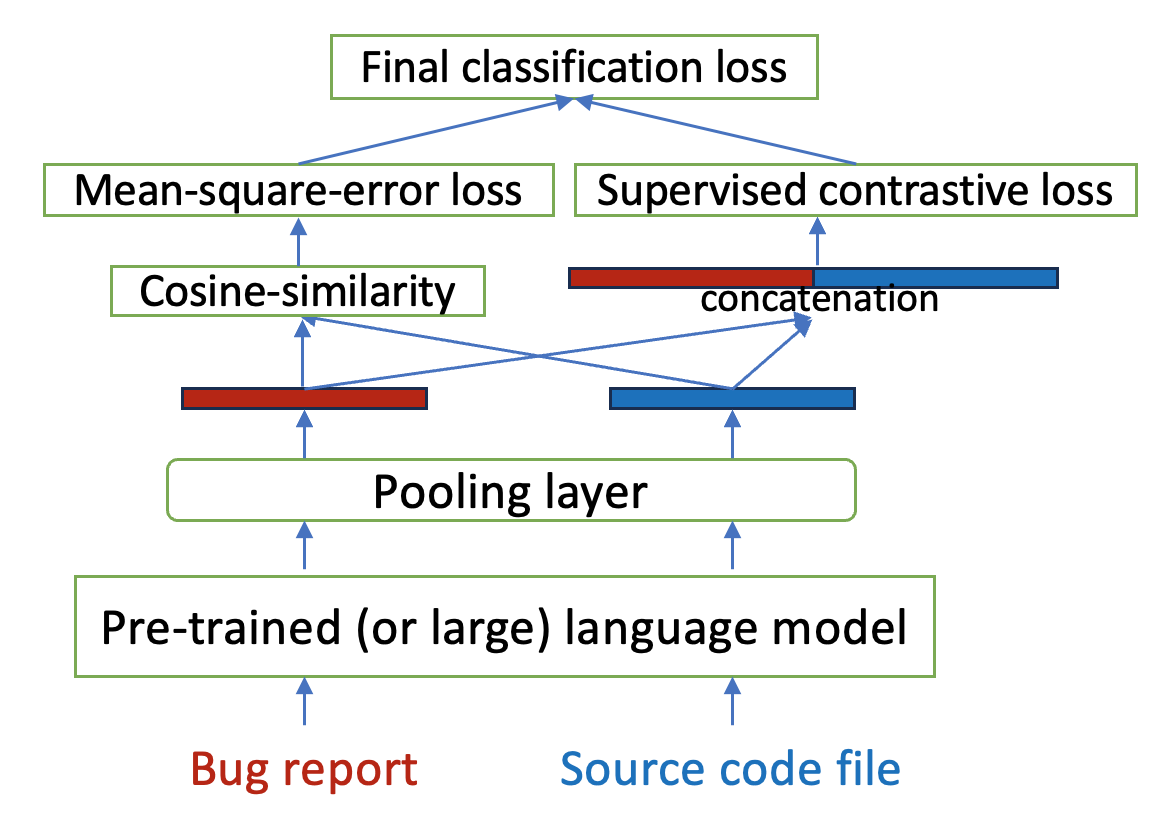} 
        \captionof{figure}{High-level fine-tuning architecture}
        \label{fig:proposed_model}
    \end{minipage}\hfill
\end{figure}

Figure~\ref{fig:proposed_model} depicts the proposed fine-tuning architecture
and Algorithm~\ref{alg:learning} summarizes the steps involved in fine-tuning
the pre-trained language model. Here, we use Supervised Contrastive Learning (SCL) \cite{khosla2020supervised} that aims to pull representations of samples from the same class closer together rather than those from different classes.
We see that using SCL can help to improve bug localization performance. These
results are explained in more detail in Section~\ref{sec:results}.

To integrate SCL during fine-tuning language models for bug localization task,
we combine the SCL loss with the mean-squared-error loss using two
hyper-parameters ($\alpha$ and $\beta$) to control these two losses, hence
learning better representations of bug reports and source code files
(Algorithm~\ref{alg:learning}, lines 8--21).

In the literature, its shown that high-quality positive and negative samples play a crucial role in the success of SCL. The conventional approach to generate negative samples is to adopt in-batch and cross-batch sampling during training \cite{wang2021syncobert}, wherein given positive sample of (bug report, source code file), negative samples can be obtained by choosing source code files from different positive samples for the bug report. 
For example, regarding the \textit{Spring-Framework} project, the bug report
about the \lstinline|ConsumesRequestCondition| class file in the package
\lstinline|org.springframework.web.reactive.result.condition| can be paired with
the class \lstinline|AbstractNameValueExpression| in the same package as a negative sample. 

This is different from our proposed technique, wherein we leverage UniXcoder~\cite{guo2022unixcoder}, state-of-the-art off-the-shelf general purpose pre-trained language model, to find similar project files to construct negative pairs.
For example, we now have the same bug report but paired with the \lstinline|ConsumesRequestCondition| class file in
the \lstinline|org.springframework.web.servlet.mvc.condition| package as a negative sample, thus helping the model to learn better representations. In particular, given a positive pair of (bug report, source
code file), we utilize UniXcoder to compute the cosine similarity scores between the embedding of the source code file and the embeddings of other project source files to obtain top $top_n$
(which we set to 10 for our experiments) most similar files. Algorithm~\ref{alg:ranking-algo-fine-tuning} shows the steps involved in the file ranking algorithm used in fine-tuning. Then, for each of ten obtained source files, we treat (bug report, obtained source file) as a negative pair. Steps involved in negative sample generation is summarized in Algorithm~\ref{alg:learning} lines 3-7.

\begin{algorithm}[!ht]
\caption{Fine-tuning pre-trained language model}
\label{alg:learning}
\DontPrintSemicolon
\SetAlgoVlined

\textbf{Input}: Positive training samples $\mathcal{P}_{training} = \{(r_{t1},
s_1, 1), (r_{t2}, s_2, 1), \cdots, (r_{tn}, s_n, 1)\}$,
project source code files $\mathcal{F} = \{f_1, f_2, \cdots, f_m\}$,
validation samples $\mathcal{V} = \{(r_{v1}, F_1), (r_{v2}, F_2), \cdots,
(r_{vl}, F_l) | F_i \subset \mathcal{F}, i \in \{ 1, 2, \cdots, l\} \}$,
model parameters \textbf{$\theta$}, an integer numeric
step $t$, the number of files to consider ($top_n$)
$\alpha=0.1$ and $\beta=0.9$.

\textbf{Output}: \textbf{$\theta$}$_{fine-tuned}$

$\mathsf{T} \leftarrow \{\}$

\For{$(r, s, 1) \in \mathcal{P}_{training}$}{

	Use UniXcoder to obtain a set $\mathsf{F} \subset \mathcal{F}$ of $top_n$ files (which are the most similar ones to $c$).
		
	\For{$f \in \mathsf{F}$}{
		$\mathsf{T} \leftarrow \mathsf{T} \cup \{(r, s, 1), (r, f, 0)\}$
	}
}
$step \leftarrow 0$

$bestAcc \leftarrow 0$

\For{epoch $\in \mathsf{range}(n\_epochs)$}{
    
    \For{batch $\in \mathsf{T}$}{
    
    	$step \leftarrow step$ + $1$
		
		$mse\_loss \leftarrow $ mean\_squared\_error\_loss($batch$)
		
		$scl\_loss \leftarrow $ supervised\_contrastive\_learning\_loss($batch$)
		
		$final\_classification\_loss \leftarrow \alpha \times mse\_loss$ + $\beta \times scl\_loss$
		
		\textbf{$\theta$} $\leftarrow$ AdamW($final\_classification\_loss$, \textbf{$\theta$})
		
		\If{(step \% t) $\mathsf{is\ equal\ to}$ 0} { 
				
			$accuracy\_at\_10 \leftarrow$ evaluation($\mathcal{F}$, $\mathcal{V}$, \textbf{$\theta$}, 10) \tcp*{Algorithm~\ref{alg:ranking-algo-fine-tuning}}
	
			\If{bestAcc $<$ accuracy\_at\_10} {
			
				$bestAcc \leftarrow accuracy\_at\_10$
				
				\textbf{$\theta$}$_{fine-tuned}$ $\leftarrow$ save\_pretrained(\textbf{$\theta$})
			}
		}
    }
}
\end{algorithm}

\begin{algorithm}[ht]
\DontPrintSemicolon
\SetAlgoVlined

\textbf{Input}: Project source files $\mathcal{F} = \{f_1, f_2, ..., f_m\}$, input reports $\mathcal{I} = \{(r_1, F_1), (r_2, F_2), ..., (r_l, F_l) | F_i \subset \mathcal{F}, i = 1, 2, ..., l \}$, an integer numeric $k$, and model parameters \textbf{$\theta$}.

\textbf{Output}: $accuracy\_at\_k$

$\mathsf{E} \leftarrow \{\}$

\For{$f \in \mathcal{F}$}{

	\textbf{e} $\leftarrow$ get\_embedding($f$, \textbf{$\theta$})
	
	$\mathsf{E} \leftarrow \mathsf{E} \cup \{\textbf{e}\}$
}
$\mathsf{R} \leftarrow \{\}$

\For{$(r, F) \in \mathcal{I}$}{

	\textbf{r} $\leftarrow$ get\_embedding($r$, \textbf{$\theta$})
	
	$\mathsf{R} \leftarrow \mathsf{R} \cup \{(\textbf{r}, F)\}$
}
$A \leftarrow \{\}$

\For{$(\textbf{r}, F) \in \mathsf{R}$}{
	
	\textbf{s} $\leftarrow$ cosine\_similarity(\textbf{r}, $\mathsf{E}$) 
	
	$ranked\_file\_list \leftarrow$ descending\_order(\textbf{s}, $\mathcal{F}$)
	
	$rank \leftarrow 0$
	
	\For{(file, score) $\in$ ranked\_file\_list[:k]}{
		
		$rank \leftarrow rank$ + $1$ 
		
		\If{$file \in F$} {
		
			\textbf{break}
		}
	}
	\eIf{$rank \leq k$} {
		
		$A \leftarrow A$ $\cup$ $\{1\}$
	}{
		$A \leftarrow A$ $\cup$ $\{0\}$
	}
} 
$accuracy\_at\_k \leftarrow$ mean($A$)

\caption{Ranking algorithm for fine-tuning.}
\label{alg:ranking-algo-fine-tuning}
\end{algorithm}

\subsection{Knowledge Distillation}
\label{sec:knowledge_distillation}
\begin{figure}[!ht]
    \centering
    \begin{minipage}{0.4\textwidth}
        \centering
        \includegraphics[width=1\textwidth]{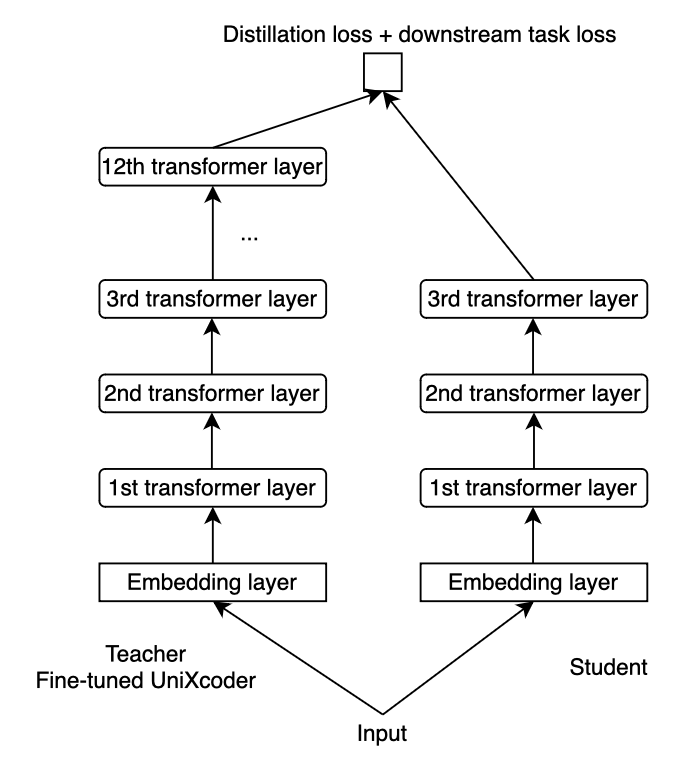} 
        \captionof{figure}{Knowledge distillation: High-level architecture.}
        \label{fig:proposed_kd_model}
    \end{minipage}\hfill
\end{figure}

In general, for task-specific distillation, given a prediction or classification downstream task (e.g., bug localization \cite{zhou2012should}), the large teacher model is obtained by fine-tuning a pre-trained language model (e.g., UniXcoder \cite{guo2022unixcoder} having 12 transformer layers). After that, the teacher model is utilized to extract the knowledge (e.g., logits or intermediate states) which are then used to guide fine-tuning the student model during distillation. Here, we assume that the teacher model is already fine-tuned and pre-defined for the given downstream task. We illustrate the overview of the model architecture in Figure \ref{fig:proposed_kd_model} and the overview of proposed fine-tuning process in Algorithm \ref{alg:kd_learning} 

To this end, we aim to answer three key questions. 
The first question is: \textit{How can we initialize the student model?}
Initializing the student model is an crucial step for improving student performance.
To address this question, we aim to distill the teacher model into a student model with fewer transformer layers (i.e., $k$ layers, k $<$ 12), wherein each student layer has the same architecture as its teacher layer (e.g., the same hidden size (i.e., $d$) and the same number (i.e., $h$) of attention heads).
This benefits from initializing the student layers directly with teacher layers.
In particular, we follow \cite{wang2023distill} to directly utilize $k$ bottom teacher layers to initialize $k$ student layers, respectively. 
In addition, we also initialize the student's embedding layer with the teacher's embedding layer.

Next, we address the second question: \textit{How can we map the
layers in the teacher model into the layers in the student model?}
We follow \cite{wang2021minilmv2} to perform last-layer knowledge distillation, i.e., only distilling the knowledge from the teacher last layer to the student last layer. 
This helps to avoid the effort of finding the best layer mapping for each downstream task.
Lastly, we address the third question: \textit{How can we transfer the knowledge from the teacher model to the student model during distillation?}
To address this question, as we aim for last-layer knowledge distillation, we consider three kinds of knowledge transfer: Prediction layer transfer, Last hidden state transfer, Last attention transfer.\\

\textit{Prediction layer transfer} minimizes the mean-squared-error (MSE) between the logits from the teacher and student models:
\begin{equation}
\mathcal{L}_{pred} = \mathsf{MSE}\left(\mathbf{z}^T, \mathbf{z}^S\right)
\end{equation}
wherein $\mathbf{z}^T$ and $\mathbf{z}^S$ denotes logits from the teacher and student, respectively.

\textit{Last hidden state transfer} minimizes the distance between the hidden states of the teacher and student last layers. We utilize MSE to measure the distance as follows:
\begin{equation}
\mathcal{L}_{hid} = \mathsf{MSE}\left(\mathbf{H}^T, \mathbf{H}^S\right)
\end{equation}
wherein $\mathbf{H}^T \in \mathbb{R}^{l\times d}$ and $\mathbf{H}^S \in \mathbb{R}^{l\times d}$ denote the hidden states of the teacher and student last layers, respectively; and $d$ is the hidden size while $l$ is the input sequence length.

\textit{Last attention transfer} aims to distill the multi-head attention matrices of the teacher last layer to the multi-head attention matrices of the student last layer.
This helps the student model to further capture semantic features and dependency knowledge encoded by attention weights of the teacher model.
We also use MSE to minimize the distance between the corresponding matrices as follows:

\begin{equation}
\mathcal{L}_{att} = \frac{1}{h} {\displaystyle \Sigma^h_{i=1}\mathsf{MSE}\left(\mathbf{A}^T_i, \mathbf{A}^S_i\right)}
\end{equation}
wherein $\mathbf{A}^T_i \in \mathbb{R}^{l\times l}$ and $\mathbf{A}^S_i \in \mathbb{R}^{l\times l}$ denote the attention matrices of the $i$-th head of the teacher and student last layers, respectively.

Finally, we follow \cite{hinton2015distilling} to use a linear combination of the downstream task and the distillation loss functions to fine-tune the student model as follows:
\begin{equation}
\mathcal{L} = \left(1 - \alpha\right)\mathcal{L}_{task} + \alpha\mathcal{L}_{distill}
\label{equa:combined_loss}
\end{equation}
wherein $\mathcal{L}_{distill} = \mathcal{L}_{pred} + \mathcal{L}_{hid} + \mathcal{L}_{att}$, and $\mathcal{L}_{task}$ denotes the downstream loss function (with respect to the student model, e.g., MSE, cross-entropy or another combination of different losses) for the given prediction/classification task. Algorithm \ref{alg:learning} summarizes the steps involved in fine-tuning the student model.

The main novelty of the proposed knowledge distillation is that we combine
current techniques into an unified framework by considering three crucial factors of knowledge distillation.
Table \ref{tab:competitors} summarizes the abilities of the current approaches
and our proposed approach for task-specific distillation. It clearly shows that none of the current approaches are, by themselves, support all the listed
features.

\begin{table*}[t]
\centering
\resizebox{0.75\textwidth}{!}{
\def\arraystretch{1.1}
\begin{tabular}{lcccccc}
\hline
{\bf Ability} & TinyBERT\cite{jiao2020tinybert} & MobileBERT \cite{2020mobilebert} & MiniLMv2 \cite{wang2021minilmv2} & \cite{wang2023distill} & Our approach \\
\hline
Initialization from bottom teacher layers & $\times$ & $\times$ & $\times$ & $\surd$ & $\surd$ \\
Last layer distillation & $\times$ & $\times$ & $\surd$ & $\times$ & $\surd$ \\
Combination of downstream task and distillation losses & $\times$ & $\times$ & $\times$ & $\times$ & $\surd$ \\
\hline
\end{tabular}
}
\caption{Comparison of proposed knowledge distillation technique with the approaches proposed in the literature.}
\label{tab:competitors}

\end{table*}

\begin{algorithm}[!ht]
\caption{Knowledge distillation process.}
\label{alg:kd_learning}
\DontPrintSemicolon
\SetAlgoVlined
\textbf{Input}: Pre-defined teacher parameters \textbf{$\theta$}$^T_{fine-tuned}$, 
training samples $\mathcal{S}_{training}$, validation samples $\mathcal{S}_{valid}$, 
a number $n$ of epochs, a number $k$ of student layers (i.e., $k=3$), and $\alpha=0.9$.

\textbf{Output}: The fine-tuned student parameters \textbf{$\theta$}$^S_{fine-tuned}$

\textbf{$\theta$}$^S$  $\leftarrow $ initialization(\textbf{$\theta$}$^T_{fine-tuned}$, $k$)  \tcp*{With respect to our first question}

$step \leftarrow 0$

$best\_result \leftarrow 0$

$t \leftarrow \lfloor \frac{n\times|\mathcal{S}_{training}|}{20} \rfloor$

\For{epoch $\in \mathsf{range}(n)$}{
    
    \For{batch $\in \mathcal{S}_{training}$}{
		
		$\mathbf{z}^S, \mathbf{H}^S, \left\{\mathbf{A}^S_i\right\}^h_{i=1} \leftarrow $ model\_forward(\textbf{$\theta$}$^S$, $batch$)
		
		$\mathbf{z}^T, \mathbf{H}^T, \left\{\mathbf{A}^T_i\right\}^h_{i=1}  \leftarrow $ model\_forward(\textbf{$\theta$}$^T_{fine-tuned}$, $batch$)
		
		$\mathcal{L}_{pred} \leftarrow \mathsf{MSE}\left(\mathbf{z}^T, \mathbf{z}^S\right)$
		
		$\mathcal{L}_{hid} \leftarrow \mathsf{MSE}\left(\mathbf{H}^T, \mathbf{H}^S\right)$
		
		$\mathcal{L}_{att} \leftarrow \frac{1}{h}\sum^h_{i=1}\mathsf{MSE}\left(\mathbf{A}^T_i, \mathbf{A}^S_i\right)$
		
		$\mathcal{L}_{distill} \leftarrow \mathcal{L}_{pred} + \mathcal{L}_{hid} + \mathcal{L}_{att}$
		
		$\mathcal{L}_{task} \leftarrow $ downstream\_loss\_function($\mathbf{z}^S$, $batch$)
		
		$\mathcal{L} \leftarrow \left(1 - \alpha\right)\mathcal{L}_{task} + \alpha\mathcal{L}_{distill}$
		 
		\textbf{$\theta$}$^S$ $\leftarrow$ AdamW($\mathcal{L}$, \textbf{$\theta$}$^S$)
		
		$step \leftarrow step$ + $1$
		
		\If{(step \% t) $\mathsf{is\ equal\ to}$ 0} { 
				
			$task\_result \leftarrow $ downstream\_evaluation(\textbf{$\theta$}$^S$, $\mathcal{S}_{valid}$) 
			
			\If{best\_result \textsf{is better than} task\_result} {
			
				$best\_result \leftarrow task\_result$
				
				\textbf{$\theta$}$^S_{fine-tuned}$ $\leftarrow$ save\_pretrained(\textbf{$\theta$}$^S$)
			}
		}
    }
}
\end{algorithm}

%% file: implementation.tex
\section{Implementation}

We fine-tune a state-of-the-art pre-trained language model (LM), UniXcoder
\cite{guo2022unixcoder}, for our bug localization task. In particular, we add a $mean$ pooling layer to the last attention layer of pre-trained LM to obtain a final embedding for each input sequence.
During fine-tuning, we use the \texttt{AdamW optimizer} \cite{loshchilov2017decoupled} to fine-tune the model by minimizing the classification loss function.

For knowledge distillation, we use the fine-tuned model as the teacher to initialize the student model; more precisely, we simply truncate 9 transformer layers from the teacher model and consider the truncated model as the student. We also add the $mean$ pooling layer to the student last layer to get a final embedding for each input sequence. Again we utilize the \texttt{AdamW optimizer} to minimize the combined loss in Equation \ref{equa:combined_loss} during fine-tuning the student model. We save the fine-tuned student model that produces the best result on the validation set and then use this fine-tuned student model to report the results on the test sets.

%% file: exprDesign.tex
\section{Experimental Design}
In this section we briefly discuss the benchmark datasets, model fine-tuning, evaluation setup and the evaluation metrics used in our experiments.
\subsection{Benchmark Datasets}

\subsubsection{Cross-language dataset}\hspace*{\fill} \label{sec:cross-lang-dataset}

For cross-language evaluation, we collected bug data from 48 real-world open-source projects written in Java, C/C++, and Golang. We leveraged the GitHub API to gather all \texttt{Closed} issues tagged as bugs (or similar terms) with associated commits and/or pull requests. The rationale is that issues marked with a \texttt{bug} tag are considered bugs by the project maintainers.

To generate the ground-truth dataset, we extracted files and methods from the commits/pull requests associated with each bug. Issues closed without an associated commit/pull request, even if marked as bugs, were excluded from our dataset. These issues lack code changes, and hence, no ground truth exists for validation. Table~\ref{tab:cross_lang_proj} summarizes the projects and bug counts.

\begin{table*}[]
\begin{tabular}{|l|l|l|l|l|}
\hline
\multicolumn{1}{|c|}{Lang} & \multicolumn{1}{c|}{\begin{tabular}[c]{@{}c@{}}\#\\ Proj.\end{tabular}} & \multicolumn{1}{c|}{\begin{tabular}[c]{@{}c@{}}\#\\ Bugs\end{tabular}} & \multicolumn{1}{c|}{Project names}                                                                                                                                                                                                                                                                                                                  & \multicolumn{1}{c|}{\begin{tabular}[c]{@{}c@{}}Dataset\\ Type\end{tabular}} \\ \hline
Java                       & 25                                                                      & 8,103                                                                  & \texttt{\begin{tabular}[c]{@{}l@{}}rocketmq, ExoPlayer, pulsar, skywalking, jsoup, spring-authorization-server, \\ logstash, selenium, junit4, guava, spring-framework, druid, dubbo, graal, \\ elasticsearch, spring-security, dbeaver, micronaut-core, visualvm, fastjson, \\ closure-compiler, gson, opengrok, junit5, spring-boot\end{tabular}} & Training                                                                    \\ \hline
C/C++                      & 10                                                                      & 2,376                                                                  & \texttt{\begin{tabular}[c]{@{}l@{}}redis, rocksdb, netdata, mbedtls, librdkafka, citus, openssl, systemd, \\ rethinkdb, timescaledb\end{tabular}}                                                                                                                                                                                                   & Testing                                                                     \\ \hline
Golang                     & 13                                                                      & 3,962                                                                  & \texttt{\begin{tabular}[c]{@{}l@{}}dagger, k9s, amass, trufflehog, meshery, k3s, nuclei, act, nats-server, \\ lazygit, go-github, go, osv-scanner\end{tabular}}                                                                                                                                                                                     & Testing                                                                     \\ \hline
\end{tabular}
\caption{Cross-language cross-project dataset}
\label{tab:cross_lang_proj}
\end{table*}

\subsubsection{Cross-project dataset}\hspace*{\fill} \label{sec:cross-proj-dataset}

We employed the Java-only dataset used by Ciborowska et al.~\cite{ciborowska2022fast} for cross-project evaluation. However, due to limitations discussed in Section~\ref{sec:motivation}, we collected the ground-truth classes and methods from the actual bug patches instead of using changeset data. A summary of these projects is presented in Table~\ref{tab:dataset_motivation}. Notably, the projects in this dataset are not part of the training dataset shown in Table~\ref{tab:cross_lang_proj}.

\subsection{Fine-tuning Configuration} 
 
We fine-tuned the off-the-shelf pre-trained language model UniXcoder~\cite{guo2022unixcoder} on the Java-only dataset (8,103 bugs) as described in Section~\ref{sec:fine-tuning}. The fine-tuning process utilized an NVIDIA Tesla A10 GPU with 24GB RAM. Similarly, the fine-tuned UniXcoder served as the teacher model for knowledge distillation, as explained in Section~\ref{sec:knowledge_distillation}. GPUs were only used for fine-tuning and knowledge distillation. Following knowledge distillation, the distilled model was employed for inference tasks. All inference reported in this paper was conducted using a CPU with 32 cores and 128 GB memory running Oracle Linux.

\subsection{Evaluation Setup}\label{sec:evaluation_setup}

As shown in~\citep{ciborowska2022fast} and~\cite{du2023pre}, the generic pre-trained language model, UniXcoder~\cite{guo2022unixcoder}, outperformed state-of-the-art bug localization techniques like Locus~\cite{wen2016locus}, BLUiR~\cite{saha2013improving}, GraphCodeBERT~\cite{graphcodebert2021} and FBL-BERT~\cite{ciborowska2022fast} without any project- or language-specific fine-tuning. Therefore, our evaluation compares the proposed approach's performance against UniXcoder. Specifically, we used the off-the-shelf UniXcoder in three configurations:

 \begin{itemize}
     \item UniXcoder$_{file}$: The baseline, off-the-shelf UniXcoder with file-level embedding generation
     \item UniXcoder$_{cs}$: Off-the-shelf UniXcoder with code segment-level embedding generation
     \item UniXcoder$_{rank}$: Off-the-shelf UniXcoder with the proposed ranking algorithm combined with commit analysis and code segment-level embedding generation. 
 \end{itemize}

\subsection{Evaluation Metrics}

\noindent\textbf{Mean Reciprocal Rank:} This measures the ability of a model to
locate the first relevant location to a bug report~\cite{voorhees1999trec}.\\

\noindent\textbf{Mean Average Precision:} This measures how well a model can
locate all locations relevant to a bug report~\cite{manning1999foundations}.\\

\noindent\textbf{Accuracy@N: } This measures the percentage of
queries that contains at least one relevant item in the first $N$
items of the retrieved list.  

%% file: results.tex
\section{Experimental Results} \label{sec:results}
We address following four research questions through our evaluation:

\subsection*{RQ1: Performance on Cross-language Dataset}

Table~\ref{tab:results_summary_cross_lang} summarizes the overall evaluation results on the cross-language dataset. In the table, 'Ours' refers to the original fine-tuned model with the proposed ranking algorithm, while 'Ours$_{distilled}$' refers to the scaled down version of the fine-tuned model after knowledge distillation. In this section, we'll focus on 'Ours' while we discuss 'Ours$_{distilled}$' in RQ4. 

Table~\ref{tab:results_summary_cross_lang} shows that the proposed approach outperforms the off-the-shelf UniXcoder pre-trained language model across all configuration settings (file, code segment, and ranking). Our approach correctly locates at least one relevant file, code segment, or commit across languages for 73-80\% of the bugs (57-67\% of the bugs) within the top 10 (top 5) results.

Individual analysis of Golang projects reveals that our approach achieved the highest accuracy at rank 10 (Acc@10) of 100\% for osv-scanner (33 bugs) and the lowest of 55\% for lazygit (40 bugs). In comparison, UniXcoder$_{rank}$ achieved the highest Acc@10 of 95.6\% for go-github (495 bugs) and the lowest of 55\% for lazygit. In our Golang dataset, the Go standard library has the largest number of bugs (2,641) for which  UniXcoder$_{rank}$ achieved an Acc@10 of 59.9\%, while our approach achieved 76.6\%, an improvement of 28

Similarly, analysis of the C/C++ projects reveals that our approach achieved the highest Acc@10 of 85.7\% for redis (49 bugs) and the lowest of 47.5\% for netdata (59 bugs). In comparison, UniXcoder$_{rank}$ achieved the highest Acc@10 of 85.8\% for librdkafka (120 bugs) and the lowest of 37.3\% for netdata. In our C/C++ dataset, systemd has the largest number of bugs (788) for which UniXcoder$_{rank}$ achieved an Acc@10 of 59\%, while our approach achieved 77\%, an improvement of 31\%. We also observe that, in general, the models perform better on the Golang dataset compared to C/C++.\\

\noindent{\textbf{Summary:} } Given a bug report, on average, there is a 73-80\% chance that at least one correct bug location can be identified within the top 10 results using the proposed approach. This, combined with the fact that our model is not trained on any of these projects or projects written in similar languages, is promising for real-world deployment.

\begin{table*}[h]
\begin{tabular}{|c|l|ll|ll|ll|ll|ll|ll|}
\hline
\multicolumn{1}{|l|}{\multirow{2}{*}{Language}} & \multicolumn{1}{c|}{\multirow{2}{*}{Model}} & \multicolumn{2}{c|}{Acc@1}                                 & \multicolumn{2}{c|}{Acc@3}                                & \multicolumn{2}{c|}{Acc@5}                                & \multicolumn{2}{c|}{Acc@10}                                & \multicolumn{2}{c|}{MAP}                                  & \multicolumn{2}{c|}{MRR}                                  \\ \cline{3-14} 
\multicolumn{1}{|l|}{}                          & \multicolumn{1}{c|}{}                       & \multicolumn{1}{c|}{Mean} & \multicolumn{1}{c|}{SD} & \multicolumn{1}{c|}{Mean} & \multicolumn{1}{c|}{SD} & \multicolumn{1}{c|}{Mean} & \multicolumn{1}{c|}{SD} & \multicolumn{1}{c|}{Mean} & \multicolumn{1}{c|}{SD} & \multicolumn{1}{c|}{Mean} & \multicolumn{1}{c|}{SD} & \multicolumn{1}{c|}{Mean} & \multicolumn{1}{c|}{SD} \\ \hline \hline
\multirow{5}{*}{C/C++}                          & UniXcoder$_{file}$                                & \multicolumn{1}{l|}{0.030}     &  \textbf{0.011}                              & \multicolumn{1}{l|}{0.048}     &   \textbf{0.017}                            & \multicolumn{1}{l|}{0.065}     &  \textbf{0.019}                             & \multicolumn{1}{l|}{0.097}     &    \textbf{0.026}                            & \multicolumn{1}{l|}{0.039}     &  \textbf{ 0.013  }                          & \multicolumn{1}{l|}{0.055}     &  \textbf{0.013}                             \\ \cline{2-14} 
& UniXcoder$_{cs}$                                 & \multicolumn{1}{l|}{0.185}     &  0.063                              & \multicolumn{1}{l|}{0.335}     &   0.097                            & \multicolumn{1}{l|}{0.41}     &  0.11                             & \multicolumn{1}{l|}{0.526}     &  0.128                             & \multicolumn{1}{l|}{0.240}     &   0.069                           & \multicolumn{1}{l|}{0.297}     &  0.077                             \\ \cline{2-14} 
                                                & UniXcoder$_{rank}$                                & \multicolumn{1}{l|}{0.260}     &  0.092                              & \multicolumn{1}{l|}{0.406}     &   0.093                            & \multicolumn{1}{l|}{0.465}     &  0.105                             & \multicolumn{1}{l|}{0.672}     &   0.128                             & \multicolumn{1}{l|}{0.289}     &   0.086                            & \multicolumn{1}{l|}{0.350}     &  0.094                             \\ \cline{2-14} 
                                                & Ours                                        & \multicolumn{1}{l|}{\textbf{0.341}}     &  0.087                              & \multicolumn{1}{l|}{\textbf{0.513}}     &   0.091                            & \multicolumn{1}{l|}{\textbf{0.576}}     &   0.101                            & \multicolumn{1}{l|}{\textbf{0.735}}     &  0.109                              & \multicolumn{1}{l|}{\textbf{0.374}}     &  0.086                             & \multicolumn{1}{l|}{\textbf{0.447}}     &      0.091                         \\ \cline{2-14}
                                                & Ours$_{distilled}$                                        & \multicolumn{1}{l|}{0.262}     &  0.083                              & \multicolumn{1}{l|}{0.426}     &   0.097                            & \multicolumn{1}{l|}{0.511}     &   0.115                            & \multicolumn{1}{l|}{0.674}     &  0.12                              & \multicolumn{1}{l|}{0.303}     &  0.081                             & \multicolumn{1}{l|}{0.371}     &      0.087                         \\ \hline \hline
\multirow{5}{*}{Golang}                             & UniXcoder$_{file}$                                 & \multicolumn{1}{l|}{0.120}     & \textbf{ 0.122}                              & \multicolumn{1}{l|}{0.192}     & \textbf{ }0.148                             & \multicolumn{1}{l|}{0.233}     & \textbf{ }0.164\textbf{ }                            & \multicolumn{1}{l|}{0.308}     &  0.183                              & \multicolumn{1}{l|}{0.129}     &  \textbf{0.100}                             & \multicolumn{1}{l|}{0.183}     &  \textbf{0.134}                             \\ \cline{2-14} 
 & UniXcoder$_{cs}$                               & \multicolumn{1}{l|}{0.277}     & 0.162                               & \multicolumn{1}{l|}{0.436}     &  0.171                             & \multicolumn{1}{l|}{0.514}     &  0.186                             & \multicolumn{1}{l|}{0.644}     &   0.199                             & \multicolumn{1}{l|}{0.301}     &    0.147                           & \multicolumn{1}{l|}{0.392}     &     0.161                          \\ \cline{2-14} 
                                                & UniXcoder$_{rank}$                                 & \multicolumn{1}{l|}{0.314}     & 0.143                               & \multicolumn{1}{l|}{0.471}     &  0.169                             & \multicolumn{1}{l|}{0.554}     &  0.157                             & \multicolumn{1}{l|}{0.705}     &   0.124                             & \multicolumn{1}{l|}{0.324}     &    0.135                           & \multicolumn{1}{l|}{0.418}     &     0.147                          \\ \cline{2-14} 
                                                & Ours                                        & \multicolumn{1}{l|}{\textbf{0.388}}     &  0.154                              & \multicolumn{1}{l|}{\textbf{0.600}}     &   0.165                            & \multicolumn{1}{l|}{\textbf{0.676}}     &   0.167                            & \multicolumn{1}{l|}{\textbf{0.803}}     &    0.122\textbf{ }                           & \multicolumn{1}{l|}{\textbf{0.392}}     &   0.156                            & \multicolumn{1}{l|}{\textbf{0.514}}     &    0.150                           \\ \cline{2-14}
                                                 & Ours$_{distilled}$                                        & \multicolumn{1}{l|}{0.347}     &  0.143                              & \multicolumn{1}{l|}{0.555}     &   \textbf{0.147}                           & \multicolumn{1}{l|}{0.612}     &   \textbf{0.155}                            & \multicolumn{1}{l|}{0.771}     &    \textbf{0.110}\textbf{ }                           & \multicolumn{1}{l|}{0.350}     &   0.135                            & \multicolumn{1}{l|}{0.469}     &    0.137                           \\ \hline
\end{tabular}
\caption{Performance of different models on cross-language dataset}
\label{tab:results_summary_cross_lang}
\end{table*}

\subsection*{RQ2: Performance on Cross-project Dataset}

Table~\ref{tab:results_summary_cross_proj} summarizes the overall evaluation results on the cross-project dataset. Since UniXcoder$_{rank}$ considerably outperformed the other two variants, UniXcoder$_{file}$ and UniXcoder$_{cs}$, in the cross-language experiment, we limited our cross-project comparison to UniXcoder$_{rank}$. Although our model is fine-tuned on a Java dataset, there is no overlap between the projects used to fine-tune our model and the projects in \textit{Cross-project dataset} (refer to Section~\ref{sec:cross-proj-dataset}).  

Table~\ref{tab:results_summary_cross_proj} shows that, similar to the cross-language evaluation, the proposed approach ('Ours') outperformed or achieved similar performance to UniXcoder$_{rank}$ at Acc@3, Acc@5, and Acc@10 across all five projects. Interestingly, UniXcoder$_{rank}$ outperformed the proposed approach at Acc@1 in the ZXing dataset, one of the smallest with only 20 bugs. However, our approach managed to achieve similar performance for Acc@3 and beyond.\\

\noindent{\textbf{Summary:}} Similar to the cross-language results, the proposed approach outperformed or achieved comparable performance for Acc@10, where there is a 55-87\% chance that at least one correct bug location can be identified within the top 10 results using the proposed approach, even when the model is not fine-tuned on the dataset it's evaluated on.

\begin{table}[]
\begin{tabular}{|c|l|l|l|l|l|l|}
\hline
Project                  & Model & \multicolumn{1}{c|}{Ac@1} & \multicolumn{1}{c|}{Ac@3} & \multicolumn{1}{c|}{Ac@5} & \multicolumn{1}{c|}{Ac@10} & \multicolumn{1}{c|}{MRR} \\ \hline \hline
\multirow{3}{*}{AspectJ} & UniX$_{rank}$  & 0.402                      & 0.462                      & 0.483                      & 0.490                      & 0.437                    \\ \cline{2-7} 
                         & Ours     & \textbf{0.507}             & \textbf{0.584}             & \textbf{0.587}             & \textbf{0.598}              & \textbf{0.544}           \\ \cline{2-7}
                         & Ours$_{dist}$     & 0.444             & 0.538             & 0.570            & 0.577              & 0.495           \\ \hline \hline
\multirow{3}{*}{JDT}     & UniX$_{rank}$  & 0.479                      & 0.596                      & 0.628                      & 0.628                       & 0.538                    \\ \cline{2-7} 
                         & Ours     & \textbf{0.628}             & \textbf{0.798}             & \textbf{0.819}             & \textbf{0.819}              & \textbf{0.709}           \\ \cline{2-7}
                          & Ours$_{dist}$     & 0.617             & 0.755             & 0.798             & 0.798              & 0.693           \\ \hline \hline
\multirow{3}{*}{PDE}     & UniX$_{rank}$  & 0.433                      & 0.517                      & 0.567                      & 0.567                       & 0.481                    \\ \cline{2-7} 
                         & Ours    & \textbf{0.700}             & \textbf{0.833}             & \textbf{0.867}             & \textbf{0.867}              & \textbf{0.763}           \\ \cline{2-7}
                         & Ours$_{dist}$    & 0.683             & 0.800             & 0.850             & 0.850              & 0.739           \\ \hline \hline
\multirow{3}{*}{SWT}     & UniX$_{rank}$  & 0.816                      & 0.847                      & \textbf{0.857}                      & \textbf{0.857}                       & 0.834                    \\ \cline{2-7} 
                         & Ours     & \textbf{0.847}             & \textbf{0.857}             & \textbf{0.857}             & \textbf{0.857}              & \textbf{0.852}           \\ \cline{2-7}
                         & Ours$_{dist}$     & 0.827             & 0.827             & 0.827             & 0.827              & 0.827           \\ \hline \hline
\multirow{3}{*}{ZXing}   & UniX$_{rank}$  & \textbf{0.500}             & \textbf{0.550}             & \textbf{0.550}             & \textbf{0.550}              & \textbf{0.525}           \\ \cline{2-7} 
                         & Ours  & 0.250                      & \textbf{0.550}             & \textbf{0.550}             & \textbf{0.550}              & 0.375                    \\ \cline{2-7}
                          & Ours$_{dist}$  & 0.150                      & 0.40             & 0.450             & 0.450              & 0.260                    \\ \hline
\end{tabular}
\caption{Performance of different models on cross-project dataset. Here Uni$_{rank}$ denotes UniXcoder$_{rank}$}
\label{tab:results_summary_cross_proj}
\end{table}

\subsection*{RQ3: Effectiveness of Proposed Fine-Tuning and Ranking Techniques}

To evaluate the effectiveness of our proposed fine-tuning and ranking techniques, we employed three variants of UniXcoder as discussed in Section~\ref{sec:evaluation_setup}. Among these, UniXcoder$_{file}$ serves as a baseline, where the embedding is generated from the entire file. As expected, Table~\ref{tab:results_summary_cross_lang} shows that embedding generation at the file level resulted in poor performance. This is because files generally contain too much information, diluting the usefulness of the embeddings. Conversely, splitting the file into chunks of 512 tokens and generating embeddings (denoted as UniXcoder$_{cs}$) for these code segments yielded better results than embedding generation at the file level. For instance, UniXcoder$_{cs}$ improves Acc@10 performance by over 400\% compared to UniXcoder$_{file}$ and by more than 100\% for C/C++ and Golang, respectively.

The variant UniXcoder$_{rank}$ combines commit analysis and code segment-level embedding generation with the proposed ranking algorithm. This approach is identical to ours except that it utilizes the off-the-shelf UniXcoder, without fine-tuning, to generate the embeddings. As shown in the table, UniXcoder$_{rank}$  significantly outperforms UniXcoder$_{cs}$ on all metrics across both C/C++ and Golang projects. This demonstrates the effectiveness of the proposed ranking algorithm. Finally, comparing the performance of our approach against UniXcoder$_{rank}$, it is clear that our fine-tuning is also very effective. For example, there is a 9\% and 14\% improvement on Acc@10 between UniXcoder$_{rank}$ and the proposed approach across C/C++ and Golang projects, respectively. \\

\noindent{\textbf{Summary:}} The proposed fine-tuning and ranking algorithms are highly effective in improving the state-of-the-art off-the-shelf pre-trained language model for bug localization.

\subsection*{RQ4: Effectiveness of Knowledge Distillation}

As discussed in Section~\ref{sec:motivation}, the inability to run complex language models on developer workstations (typically CPUs) is a major hurdle for wider adoption of these bug localization techniques. To address this challenge, we propose a knowledge distillation technique that aims to achieve performance comparable to the original model while allowing the model to run on a CPU. The results achieved by the proposed distilled model are presented in Table~\ref{tab:results_summary_cross_lang} (as Ours$_{distilled}$) and Table~\ref{tab:results_summary_cross_proj} (as Ours$_{dist}$). The results indicate that the proposed distillation technique achieves comparable performance on both cross-language and cross-project datasets. For example, comparing the original and the distilled models, the performance reduction on Acc@10 is 8.2\% and 4\% for C/C++ and Golang projects, respectively. This trend holds true for the cross-project dataset as well. Importantly, the proposed distilled model outperforms UniXcoder$_{rank}$ on all benchmark datasets and metrics, except for Acc@1 on ZXing.

Furthermore, Table~\ref{tab:kd_time} shows that the proposed distilled model takes nearly 4 times less time to generate an embedding compared to the original model on a CPU. It takes around 1.6 seconds to generate an embedding using the original model, while the distilled model takes only 0.4 seconds. While 1.6 seconds might seem reasonable at first glance, this time quickly adds up when dealing with a large project containing hundreds of thousands of code segments. \\

\noindent{\textbf{Summary:}} The proposed distillation technique achieves performance comparable to the original model while reducing embedding generation time by 4 times.

\begin{table}[ht]
\begin{tabular}{|c|c|}
\hline
Model & Avg. embedding generation time \\ \hline
Ours$_{original}$  & 1550ms/sample                                                                    \\ \hline
Ours$_{distilled}$  & 390ms/sample                                                                     \\ \hline
\end{tabular}
\caption{Embedding generation time on a CPU}
\label{tab:kd_time}
\end{table}

%% file: threats.tex
\section{Threats to Validity}

The findings presented in this paper are susceptible to several threats to validity. A key threat lies in the generation of the ground-truth dataset for cross-language evaluation (Section~\ref{sec:cross-lang-dataset}). While we relied on GitHub issue tags to identify bugs, we did not manually curate them to confirm their validity. Additionally, we leveraged commits and pull requests referenced in the GitHub \texttt{CloseEvent} to identify files and code segments relevant to the bug (i.e., ground-truth generation). These referenced commits might not always be relevant; however, our limited manual analysis of \texttt{CloseEvent} suggests they are indeed relevant.

Furthermore, we fine-tuned our model using the state-of-the-art pre-trained language model, UniXcoder~\cite{guo2022unixcoder}. We took every precaution to ensure no overlap between the fine-tuning dataset and our test set to avoid data leakage. However, as UniXcoder is a generic language model trained on a vast amount of source code and documentation available on GitHub, there is a possibility that the bug reports in our test dataset were previously encountered by UniXcoder. Nonetheless, we argue that UniXcoder's generic nature, meaning it is not specifically trained for the downstream task of bug localization, makes its use acceptable in our scenario, as evidenced by similar pre-trained language model based techniques employed in prior literature~\citep{ciborowska2022fast, du2023pre}. To completely eliminate this threat, one would need to train a language model from scratch with no exposure to the test dataset, which is impractical in most cases.

%% file: relatedWork.tex
\section{Related Work}\label{sec:relatedWork}

Bug localization is a well-established field with a wealth of research documented in the literature. Knowledge distillation, on the other hand, is an emerging research area. In this section, we will briefly discuss relevant bug localization and knowledge distillation work related to our research

\subsection{Bug Localization}

Many approaches have been proposed for bug localization.
Early approaches \cite{lukins2010bug,zhou2012should,wen2016locus} focus on information retrieval (IR)-based bug localization. In general, both the bug reports and the source code files are represented as weighted token vectors using token frequency and inverse document frequency (tf-idf) \cite{salton1975vector}. They can also be represented as probability distribution vectors over latent topics using LDA topic model \cite{blei2003latent}.
Then, the similarity between the bug report and the source files is measured as the cosine similarity between their vectors. Finally, the source code files are ranked based on their similarities to the bug report.

Recently, as pre-trained LMs have become an important trend for code representation learning, some up-to-date approaches for bug localization \cite{zhu2022bl,liang2022modeling,ciborowska2022fast,chakraborty2023rlocator}  leverage pre-trained LMs (e.g., CodeBERT \cite{codebert2020}) to obtain the vector embeddings of the bug reports and the source code files. The obtained results show that these pre-trained language model-based approaches outperform the IR-based approaches~\citep{kim2013should,lukins2010bug,moreno2014use, saha2013improving, wang2014version, wong2014boosting, zhou2012should}.

As discussed in Section~\ref{sec:motivation}, while recent changeset-based bug localization techniques~\citep{ciborowska2022fast,du2023pre, wen2016locus} show promise in theory, their real-world effectiveness remains unclear. This stems from the fact that bug-related changesets aren't always small and manageable. In practice, many bugs arise from newly introduced features~\cite{catolino2019not}, resulting in larger changesets that involve committing entire files.

\subsection{Knowledge Distillation}

DistillBERT \cite{sanh2019distilbert} is initialized with the same general architecture as the teacher BERT \cite{devlin2018bert}, but taking one layer out of two from BERT. 
DistillBERT is then pre-trained using a linear combination of the distillation loss (over the soft target probabilities of the teacher BERT) and the masked language modeling loss. After that, DistillBERT is further fine-tuned for a given downstream task.

TinyBERT \cite{jiao2020tinybert} use an uniform layer-to-layer distillation mapping to transfer hidden states and self-attention weight matrices from the teacher layers to the student layers. 
MobileBERT \cite{2020mobilebert} is designed to be as deep as BERT$_{LARGE}$ while each layer is made much narrower via using bottleneck structures to keep the hidden size as same as the teacher's hidden size. 

MiniLMv2 \cite{wang2021minilmv2} extends self-attention distillation in MiniLM \cite{wang2020minilm} by employing multi-head self-attention relations to have more fine-grained self-attention knowledge to train the student. 

\cite{wang2023distill} follows TinyBERT to perform intermediate layer and prediction layer distillation to transfer knowledge from a fine-tuned RoBERTa$_{BASE}$ \cite{liu2019roberta} into a 3-layer student model for each GLUE task, wherein the student learns from the same teacher’s layers that were used for initialising the student when performing intermediate layer distillation. 

%% file: conclusion.tex
\section{Conclusion}
This paper presented a novel pre-trained language model based technique that tackles the challenge of bug localization across project and language boundaries. Our approach leverages contrastive learning and a unique ranking method to locate bugs at various granularities.  Furthermore, the model excels in generalizability, effectively identifying bugs in unseen codebases without project-specific training.  To address practical deployment, we propose a CPU-compatible solution with an efficient knowledge distillation technique.  Our results demonstrate a high probability (73-80\% for cross-language and 55-87\% for cross-project) of identifying correct bug locations within the top 10 results, even without fine-tuning the model on the specific datasets evaluated. This work offers a promising avenue for practical and efficient bug localization in real-world development scenarios.